\documentstyle[aps,eqsecnum,version2]{revtex}
\begin{document}
\begin{title}
Parametric Spectral Correlations of Disordered Systems in
 the  Fourier Domain
\end{title}
\author{ Italo Guarneri$^{1,2}$,
 Karol \.Zyczkowski$^{1,3}$, Jakub Zakrzewski$^{3,4}$,
Luca Molinari$^5$,\\
and Giulio Casati$^{1,6}$}
\begin{instit}{ $^1$Universita' di Milano-sede di Como- via Lucini 3,
22100 COMO, Italy\\

$^2$INFN, Sezione di Pavia, Via Bassi 6, 27100 Pavia, Italy.\\

$^3$Instytut Fizyki Uniwersytetu  Jagiello\'nskiego, ul. Reymonta 4,
 30-059 Krak\'ow, Poland\\

$^4$Laboratoire Kastler Brossel, Universit\'e
Pierre et Marie Curie,\\  T12, E1,
4 place Jussieu, 75272 Paris Cedex 05, France\\

$^5$Dipartimento di Fisica, Universita' di Milano, Via Celoria 16,
Milano, Italy}\\

$^6$INFN, Sezione di Milano, via Celoria 16, Milano, Italy.
\end{instit}\date{\today}

\begin{abstract}
A Fourier analysis of parametric level dynamics for random matrices
periodically depending on a phase is developed. We demonstrate both
theoretically and numerically that under very general conditions the
correlation $
C(\varphi )$ of level velocities is singular at $\varphi =0$ for any
symmetry class; the singularity is revealed by algebraic tails in Fourier
transforms, and is milder, the stronger the level repulsion in the chosen
ensemble. The singularity is strictly connected with the divergence of the
2nd moments of level derivatives of appropriate order, and its type
 is specified to leading terms for Gaussian,
stationary ensembles of GOE, GUE, GSE types, and for the Gaussian ensemble
of Periodic Banded Random Matrices, in which a breaking of symmetry occurs.
In the latter case, we examine the behaviour of correlations in the
diffusive regime and in the localized one as well, finding a singularity
like that of pure GUE cases. In all the considered ensembles we study the
statistics of the Fourier coefficients of eigenvalues, which are Gaussian
distributed for low harmonics, but not for high ones, and the
distribution of kinetic energies.
\end{abstract}

\draft


\section{Introduction}

In this paper we shall be concerned with spectral properties of random
matrices $H(\varphi )$, periodically depending on a phase $\varphi \in
\left[ 0,2\pi \right] $. Random matrices \cite{mehta} are often used to
describe statistical properties of disordered mesoscopic systems \cite
{GorEli,Efetov,AltShkl}. In particular, random matrices parametrically
dependent on a phase are used as models for conduction in small metallic
rings threaded by an Aharonov-Bohm flux $\varphi h/e$, and great attention
is being devoted to the dependence of their eigenvalues $e_i(\varphi )$ on
the phase $\varphi $ \cite{SzA93,Ben93,Alts,PBRM,anders,BrMont}. Objects of
particular interest are the statistics of level slopes $e_i^{\prime
}(\varphi )$, (also called level velocities; a prime denotes a derivative
with respect to $\varphi $), of level curvatures $e_i^{\prime \prime
}(\varphi )$, and velocity-velocity autocorrelations

\begin{equation}
\label{corr}C(\varphi ):=\frac 1{2\pi \Delta ^2}\int\limits_0^{2\pi
}\left\langle e_i^{\prime }(\varphi ^{\prime })e_i^{\prime }(\varphi
^{\prime }+\varphi )\right\rangle d\varphi ^{\prime }=\frac 1{\Delta
^2}\left\langle \overline{e_i^{\prime }(\varphi ^{\prime })e_i^{\prime
}(\varphi ^{\prime }+\varphi )}\right\rangle ,
\end{equation}
where $\left\langle \text{ }\right\rangle $ denotes ensemble averaging, the
bar denotes phase averaging, and $\Delta $ is the average level spacing.

Autocorrelation functions that are closely related to (\ref{corr}) (at least
insofar as a single-particle description is valid) are also accessible to
laboratory experiments on quantum dots\cite{Siv}.

To the extent that random matrices can be taken as models for the
Hamiltonians of quantum systems that become chaotic in the classical limit,
the identification of ''universal'' properties of parametric level dynamics
is also relevant for quantum chaology \cite{fritz}. In that case the phase $%
\varphi $ represents any external parameter which determines the spectrum.
The motion of levels as functions of $\varphi $ may be thought of as a
dynamics of fictitious particles (with positions $e_i$), with $\varphi $
playing the role of ``time''.

In a series of papers, Altshuler and coworkers \cite{Alts} have shown that
for disordered metallic systems the parametric statistics are in fact
universal, after rescaling the levels to unit mean spacing and also
rescaling the parameter $\varphi $ as $X=\pi \varphi \sqrt{c}$, with $c=C(0)$
denoting the mean squared velocity of energy levels. Therefore, the only two
system -specific parameters are $\Delta $ and $c$; the latter has the
meaning of twice the average kinetic energy of the level moving in the time $%
\varphi $. An example of such a parametric universality has been discussed
earlier in studies of level curvatures \cite{grn,saher,ZD93}.

In particular, the theory \cite{Alts} leads to a remarkable scaling property
of the velocity-velocity autocorrelation (\ref{corr}):

\begin{equation}
\label{scal}C(\varphi )\approx c\Phi (\pi \varphi \sqrt{c})=c\Phi (X).
\end{equation}
In Eq.(\ref{scal}) $\Phi (.)$ is an universal function, only dependent on
the general symmetry properties of the chosen matrix ensembles. Determining
the shape of this universal function is manifestly an important task. At
large $X$, the behaviour $\Phi (X)\sim -X^{-2\text{ }}$ was theoretically
predicted \cite{Alts,SzA93,Ben93} and numerically tested \cite{jz}; instead,
the small $X$ behaviour appear to critically depend on the chosen ensemble.
Explicit expressions have been obtained \cite{Alts} for a closely related
[but distinct from $C(\varphi )$] autocorrelation function at fixed energy.
The resulting expressions are valid for homogeneous ensembles only, in which
the symmetry class of the random matrix is independent of $\varphi $, and
are quite complicated (for the orthogonal ensemble a triple integral with
regularization is required). A global approximation for $\Phi (X)$ has been
proposed \cite{jz}. For the case of a classically chaotic system subject to
a Aharonov-Bohm flux Berry and Keating \cite{BK94} have recently obtained a
semiclassical approximation for $C(\varphi )$ having the form of an
everywhere analytic function of $\varphi $.

In this paper we shall investigate correlations (\ref{corr}) in various
ensembles of phase--dependent random matrices, using Fourier analysis, which
is particularly well suited to this purpose, because of periodicity, and
also because the singularities of $C(\varphi )$ are easily detected by this
technique. We shall consider the case of homogeneous Gaussian ensembles,
with orthogonal (GOE), unitary (GUE) and symplectic (GSE) symmetry, and also
the physically important situation, in which the time - reversal symmetry is
broken and a transition from GOE to GUE occurs while switching on the
''Aharonov--Bohm flux'' $\varphi $. The latter case will be analyzed on the
ensemble of Periodic Band Random Matrices (PBRM)\cite{PBRM}, which exhibits
the three characteristic regimes of conduction in disordered solids, namely,
the localized, the (proper) diffusive, and the ballistic one.

A general theoretical approach to the Fourier analysis of spectral
correlations is
developed in Section II, where it is shown that, under very general
conditions, $C(\varphi )$ must have a singularity at $\varphi =0$, signalled
by an algebraic tail in its Fourier transform. This result is valid even in
cases (such as homogeneous GUE, GSE) which were hitherto believed to be
singularity-free. The singularity is however milder, the stronger the level
repulsion in the considered ensemble. We determine its exact type for
various symmetry classes. For the homogeneous cases our predictions are
confirmed by numerical results, presented in Section III, and by the exact
solution for matrices of rank $2$, presented in the Appendix. The model of
Periodic Band Random Matrices corresponding to the Aharonov - Bohm case is
analyzed in Section IV. We present theoretical arguments and numerical
results supporting the thesis that for this case the behaviour of the
correlation function at $\varphi =0$ is similar to the one observed in the
homogeneous GUE case, and we compare our data to other theoretical
predictions. Furthermore we present distributions of avoided crossings for
PBRM, which are deeply related to the properties of the velocity correlation
function, and briefly discuss the modifications of the correlation function
which occur on moving towards the localized regime.

Further applications of Fourier analysis presented in this paper include the
statistics of various quantities. Among these, the Fourier amplitudes of
level velocities (currents), which are found to be Gaussian for low
harmonics but not for high ones, and the mean squared current, which appears
to follow a $\chi ^2$ distribution with an appropriate number of degrees of
freedom. Finally, we discuss the moments of level derivatives of order $k$
with respect to phase $\varphi $, and find them either to diverge (if $k$ is
not small enough) or to be directly related, via the scaling law, to the
average kinetic energy of levels.

\section{Fourier Analysis.}

In full generality, by a Periodic Random Matrix (PRM) we mean a random
Matrix $H(\varphi )$ periodically depending on a phase $\varphi $; more
precisely, a matrix-valued, periodic, stochastic process, taking values in a
class of matrices having a definite symmetry (symmetric, complex-hermitean,
symplectic), which is not necessarily the same for all values of $\varphi $.
An ensemble of PRM is not an ensemble of matrices, but an ensemble of {\it %
trajectories} $H(\varphi )$ in a space of matrices; every such trajectory
will be called a realization of the ensemble. In this paper we shall consider
stationary ensembles, such that their statistics is invariant under
translations $mod(2\pi)$ of the phase $\varphi$, and non-stationary ones as
well.
Although the numerical results
described in this paper were obtained for some very specific, well-defined
ensembles, the argument to be presently described is applicable under rather
general assumptions, hence to a much broader class of ensembles.

Our first assumption is that, with probability one, the trajectory $%
H(\varphi )$ is an entire function of $\varphi $. On the strength of this
assumption, the eigenvalues $e(\varphi )$ of almost any realization $%
H(\varphi )$ are analytic functions of $\varphi $, and their behaviour in a
neighborhood of any point $\varphi $ in $[0,2\pi ]$ can be described by
analytic perturbation theory. The corresponding convergence radii, however,
are realization-dependent. If they can become arbitrarily small in a
statistically significant fraction of the ensemble, perturbative methods
will not be applicable to the calculation of certain statistical averages.
This is precisely the situation we are going to discuss below.

Given a realization $H(\varphi)$, any eigenvalue $e(\varphi)$ can be
expanded in a Fourier series. The Fourier expansions we shall use are :

\begin{equation}
\label{four}e(\varphi )=\sum\limits_{-\infty }^{+\infty }a_ne^{in\varphi
}\qquad C(\varphi )=\sum\limits_{-\infty }^{+\infty }c_ne^{in\varphi },
\end{equation}
where $a_n=a_{-n}^{*}$ and, on account of Eq.(\ref{corr}), $c_n=n^2\Delta
^{-2}\left\langle \left| a_n\right| ^2\right\rangle .$ The statistical
average $\left\langle .\right\rangle $ is meant as an average over all the
eigenvalues lying in a selected energy range (in the following we drop a
subscript $i$ denoting a given level in $e_i$), followed by an average over
all realizations in some ensemble. Because $C(\varphi )$ is a correlation
function, it must be even, $C(\varphi )=C(2\pi -\varphi )$. From (\ref{four}%
) it follows that:

$$
c=C(0)=\sum\limits_{-\infty }^{+\infty }c_n.
$$
If, in the chosen ensemble, a scaling law of the form (\ref{scal}) holds,
then, at $c>>1$ :

\begin{equation}
\label{scal1}c_n\approx \sqrt{c}P(\frac n{\sqrt{c}}),
\end{equation}
with $P(t)$ a function related to the Fourier transform of $\Phi (X)$ [Eq.(%
\ref{scal})]:
\begin{equation}
\label{Pt}P(t)=(2\pi ^2)^{-1}\int \Phi (X)e^{-itX/\pi }dX
\end{equation}
It follows that $P(t)$ defines a probability distribution on the real line,
with (not necessarily finite) moments $A^{(\alpha )}=\langle |t^\alpha
|\rangle $. In particular, the width of the Fourier spectrum is measured by:
\begin{equation}
\label{scal2}\langle |n|\rangle :=\frac{\sum\limits_{-\infty }^{+\infty
}|n|c_n}{\sum\limits_{-\infty }^{+\infty }c_n}\approx \sqrt{c}\int \left|
t\right| P(t)dt=A^{(1)}\sqrt{c}
\end{equation}
Even moments are related in an obvious way to derivatives of $\Phi $ at
the origin:
\begin{equation}
\label{mom}\Phi^{(2k)}(0)=(-1)^k\pi^{-2k}A^{(2k)}
\end{equation}
 In ensembles for which the scaling theory\cite{Alts} is valid,
the functions $\Phi ,P$ and the constants $A^{(\alpha )}$ depend on the
universality class only, which will be occasionally specified in the
following by a suffix $\beta $ (as, e.g.,in $\Phi _\beta $), with $\beta
=1,2,4$ for the GOE, GUE, GSE universality classes, respectively; or by the
suffix $AB$ in the symmetry-breaking, GOE-GUE case. Eqs.(\ref{scal1}),(\ref
{scal2}) provide a way to check the scaling hypothesis in the Fourier
domain. For the specific ensembles considered in the following Sections they
were indeed confirmed by our numerical computations, to be discussed later.

Following our previous assumption, $H(\varphi )$ can be analytically
continued to the whole complex $\varphi -$plane. Any eigenvalue $e(\varphi )$
is an analytic function of the complex variable $\varphi ,$ except for
branch-points (BP) $z_j=\varphi _{0j}+i\gamma _{j\text{ }}$corresponding to
complex level crossings\cite{Heiss}. The index $j$ labels the BPs,
$j=1,...{\cal N},$
with ${\cal N}$ the total number of BPs exhibited by the given level as a
function of $\varphi $ in the complex domain. Since $e(\varphi )$ is real at
real $\varphi ,$ these BPs are distributed symmetrically with respect to the
real axis, but none of them falls on the real axis itself
\footnote{Although crossing of different
 eigenvalues at real values of $\varphi$ is possible,
such real crossings would not spoil the analiticity of $e(\varphi)$.}.
Therefore, given a realization $H(\varphi )$, the function $e(\varphi )$ is
analytic inside a strip of halfwidth $\Gamma $ around the real axis, $\Gamma
$ being the smallest of the $\left| \gamma _j\right| $. In order to estimate
the complex Fourier coefficients

$$
a_n=\frac 1{2\pi }\int\limits_0^{2\pi }e^{-in\varphi }e(\varphi )d\varphi,
$$
with $n>0$ we use the multiple path illustrated in Fig.\ref{Fig.1}. In the
following we assume $n>0$ (if $n<0$ the path must be reflected with respect
to the real axis; the corresponding analysis differs from the following one
by obvious modifications). In this way

\begin{equation}
\label{a}a_n=\sum_{j=1}^{{\cal N}} I_j,
\end{equation}
where $I_j$ is the complex contribution of the $j-$th branch-line:

\begin{equation}
\label{int1}I_j=\frac{e^{-inz_j}}{2\pi }\int\limits_0^\infty dx\
e^{-nx}\delta e_j(x),
\end{equation}
and $\delta e_j(x)$ is the jump of $e(\varphi )$ across the $j-$th branch
cut, at the point $\varphi =z_j+ix$. The behaviour of the Laplace integral (%
\ref{int1}) at large $n$ is determined by the behaviour of $\delta e_j(x)$
close to the BP, as we shall presently discuss. We concentrate on BPs with
$\gamma _j<<1,$ whose contribution will dominate at large $n$. We shall
consider simple BPs only; in other words, we shall attribute zero
statistical weight to crossings involving more than two levels. If a
realization $H(\varphi )$ ($\varphi $ complex) is thought of as a two-(real)
parameters family of matrices, then generically such a family will not
contain matrices exhibiting higher-order degeneracies, because
multiple level crossings cannot generically be achieved by varying just two
real parameters.
Therefore, the assumption we
need to justify our neglecting higher-order BPs is just that such
non-generic
realizations $H(\varphi)$ have zero probability; this is a smoothness
assumption on the statistical distribution of the ensemble, which
appears fairly general.

For $\varphi $ real and close to $\varphi _{0j\text{ }}$ the considered
level is undergoing a narrow avoided crossing. It is in fact close to
degenerate with a different level, the two eigenvalues being approximately:

\begin{equation}
\label{split}e_{\pm }(\varphi )\approx E_j\pm h_j\sqrt{\gamma _j^2+(\varphi
-\varphi _{0j})^2}.
\end{equation}
Carrying this to complex $\varphi $ we get, for $\varphi $ close to $z_j,$ $%
\delta e_j(\varphi )\approx 2h_j\sqrt{\gamma _jx}$. Substituting into (\ref
{int1}), and using known estimates for Laplace integrals \cite{Italo1}, we
obtain

\begin{equation}
\label{int2}I_j\sim \frac{h_j}{\pi} \sqrt{\gamma _j}n^{-3/2}e^{-inz_j}.
\end{equation}
We now introduce the single-level velocity correlation function (CF):

\begin{equation}
\label{RR} R(\varphi ):=\Delta ^{-2}\overline{e^{\prime }(\varphi ^{\prime
})e^{\prime }(\varphi ^{\prime }+\varphi )}=\sum\limits_{n=-\infty
}^{+\infty }r_ne^{in\varphi },
\end{equation}
(the bar denoting average over $\varphi^{\prime}$),
which depends on the chosen eigenvalue, and on the realization as well. This
function is connected to the original CF by $C(\varphi )=\left\langle
R(\varphi )\right\rangle $, and $c_n=\left\langle r_n\right\rangle .$ From (%
\ref{int2}) we obtain:

\begin{equation}
\label{rfi}r_n=\Delta ^{-2}n^2\left| a_n\right| ^2\sim \frac \Gamma
ne^{-2n\Gamma }.
\end{equation}
Thus, prior to ensemble averaging $R(\varphi )$ has an exponentially
decaying Fourier expansion, and is therefore analytic in a strip of
halfwidth $2\Gamma $. From (\ref{rfi}) it actually follows that $R(\varphi )$
has logarithmic singularities at $\varphi =\pm 2i\Gamma +2k\pi ,$ with
leading terms $\Gamma \log (\varphi \pm 2i\Gamma +2k\pi )$. Therefore, for
real $\varphi $ close to 0, $R(\varphi )$ behaves like $\Gamma \log (\varphi
^2+4\Gamma ^2).$ On averaging over the matrix ensemble, arbitrarily small $%
\Gamma $ come into play, that is, the logarithmic singularities of $%
R(\varphi )$ come arbitrarily close to the real axis. As a result, $%
C(\varphi )$ will not be analytic at $\varphi =0$ any more; the type of
singularity will depend on the statistical weight which is given to small $%
\Gamma $ in the chosen ensemble. Actually, a standard probability-theoretic
argument, quite independent of the above elaborations, shows that in order
for $C(\varphi )$ to be analytic, too small $\Gamma $ must not be allowed at
all!

In fact, if we assume $C(\varphi )$ to be analytic, then, being also
periodic, it must have an analyticity strip of (strictly) nonzero width $2L>0
$. Then the coefficients $c_n$ must satisfy
$$
c_n<K_{L^{\prime }}e^{-2L^{\prime }|n|},
$$
for all $L^{\prime }<L$ (with appropriate constants $K_{L^{\prime }}$).
Since $c_n=n^2\Delta^{-2}\langle |a_n|^2\rangle $, the probability
$p_n(L^{\prime
\prime })$ that $|a_n|$ be larger than $e^{-L^{\prime \prime }|n|}$ will
satisfy the Chebyshev inequality \cite{Italo2}:
$$
p_n(L^{\prime \prime })<c_nn^{-2}\Delta^{2}e^{2L^{\prime \prime }|n|}
<K_{L^{\prime
}}\Delta^{2}e^{-2(L^{\prime }-L^{\prime \prime })|n|}.
$$
Therefore $\sum_np_n(L^{\prime \prime })<+\infty $ as soon as $L^{\prime
\prime }<L$. Then the Borel-Cantelli Lemma of Probability theory \cite{Borel}
ensures that, with probability one, all but finitely many of the
coefficients $a_n$  satisfy $|a_n|<e^{-L^{\prime \prime }|n|}$ for any $%
L^{\prime \prime }<L$. In turn, this implies that $e(\varphi )$ is analytic
in a strip of width $L$ with probability one; that is, the probability of a
branch-point lying closer than $L$ to the real axis is zero. Therefore, in
order to get an everywhere analytic $C(\varphi )$, the ensemble must be such
that the distribution of the imaginary part of BPs has a finite gap. Since
the latter distribution is clearly related to the distribution of the
widths of avoided crossings, this appears to be rather an exceptional
situation.

In order to get more precise information about the nature of the
singularities of $C(\varphi)$, a knowledge of the statistical distribution
of the BPs is needed, which cannot be obtained under the general assumptions
made up to now. We shall then assume that this distribution is described by
a density $p(\gamma ,\varphi _0),$ and also that different BPs give
uncorrelated contributions. Furthermore, in the rest of this Section we
shall restrict ourselves to stationary ensembles, such that all statistical
properties are independent of $\varphi$; explicit examples will be given in
the next Section. In such cases we can assume statistical independence of $%
\varphi _0$ and $\gamma $, and a uniform distribution for the former. In
other words, $p(\gamma ,\varphi _0)=f(\gamma )/(2\pi).$ Then, from Eqs.(\ref
{a}) and (\ref{int2}) we obtain

\begin{equation}
\label{c}c_n\sim \frac {\left\langle h^2\right\rangle {\cal N}} { \pi
\Delta^2 n } \int\limits_0^\infty d\gamma \ \gamma e^{-2n\gamma }f(\gamma ).
\end{equation}
In order to assess how does this expression scale with $c,$ we first notice
from (\ref{split}) that the quantity $h$ sets the scale for the level
velocity close to the avoided crossing. Therefore we assume $\left\langle
h^2/\Delta ^{2}\right\rangle $ to scale as the variance of the rescaled level
velocity, that is, as $c=C(0)=\sum c_n.$ At the same time, $h\gamma $ is the
width of the avoided crossing, and this identifies the scale of $\gamma $
with that of $\Delta /h\sim c^{-1/2}.$ On such grounds, setting $\gamma
=xc^{-1/2},$ we expect

$$
f(\gamma )=c^{1/2}F(\gamma c^{1/2}),
$$
with $F(x)$ a scale-independent distribution. Substituting this into (\ref{c}%
) we obtain

\begin{equation}
\label{c1}c_n\sim \alpha {\cal N} \frac{\sqrt{c}}n\int\limits_0^\infty
dxF(x)xe^{-2xn/\sqrt{c}},
\end{equation}
(in the rest of this Section, $\alpha $ will denote undetermined numerical
factors).  In order to estimate $
{\cal N} ,$ we identify its order to magnitude with that of the number of
avoided crossings experienced by $e(\varphi )$ as $
\varphi $ varies between $0$ and $2\pi,$ which has been found to be
proportional to $\sqrt{c}$  using two different approaches\cite
{wilk,ZDK93}.  We are thus
finally led to

\begin{equation}
\label{fin}c_n\sim \alpha \frac c n\int\limits_0^\infty dxF(x)xe^{-2xn/\sqrt{c%
}}.
\end{equation}
This formula has a number of implications. In the first place, it is exactly
of the form (\ref{scal1}); the above argument has led us to an independent
confirmation of the scaling law at $n>>\sqrt{c.}$ Second, it yields an
asymptotic estimate for the universal function $P(t)$ at large $t:$

\begin{equation}
\label{fini} P(t)\sim \frac \alpha t\int\limits_0^\infty dxF(x)xe^{-2tx}.
\end{equation}
It does not appear legitimate, however, to use formula (\ref{fin}) outside
the asymptotic regime $n>>\sqrt{c}$, because of the underlying asymptotic
estimate (\ref{int2}).

To proceed further, we need to specify $F(x)$, or at least its behaviour at $%
x<<1,$ which determines the asymptotics of (\ref{fin}) at large $n.$ The
variable $x$ is, apart from rescaling, the same as $\gamma $, which is in
turn closely related to the width of an avoided crossing. For the case of
Gaussian ensembles, the statistics of the latter quantity has been studied
in \cite{wilk,ZK91,ZDK93}, where the probability of small avoided crossings
of size $\epsilon $ was found to scale as $\epsilon ^{\beta -1},$ with $%
\beta $ the repulsion parameter characteristic of the given ensemble. This
result is actually very general. According to (\ref{split}), the probability
of a small level spacing less than $\epsilon $, which scales as $%
\epsilon^{\beta+1}$, is also proportional to the probability of getting a BP
inside a small circle of radius $\epsilon $ in the complex $\varphi -$plane.
The latter probability scales as $\epsilon \int_0^\epsilon fd\gamma $,
whence $f(\epsilon )\sim \epsilon ^{\beta -1}.$

Our last assumption will be that level repulsion in our (stationary)
ensemble is ruled by a repulsion exponent $\beta$. Then, substituting $%
F(x)\sim x^{\beta-1}$ ( at small $x$) into Eq.(\ref{fini}) we finally get
\begin{equation}
\label{tail}P(t)\sim \alpha t^{-2-\beta }
\end{equation}
which is the main result of this Section. It implies that{\it \ in no
stationary ensemble can $C_\beta (\varphi )$ be a smooth function at $%
\varphi =0,$} because its ($\beta +1)$-th derivative must have a singularity
there. It should be noted that  the asymptotic behaviour
$c_n\sim n^{-2-\beta}$, which leads to this conclusion,
directly follows from eqn.(\ref{c}) and from the
fact that the probability of small $\gamma$ scales as $\gamma^{\beta-1}$,
independently of the scaling considerations following eqn.(\ref{c}).

The nature of the singularity associated with the
behaviour (\ref{tail}) now follows, either from known results in Fourier
analysis, or, more directly, from statistically averaging the previously
described small-$\varphi $ behaviour of $R(\varphi )$. The resulting
small-$\varphi$
expansions for the cases $\beta=1,2,4$, written in terms of the scaled
variable $X=\pi\varphi \sqrt{c}$, are (up to leading terms only):

\begin{equation}
\label{singr}
\begin{array}{c}
C_1(\varphi )/c\sim 1+ b_1 X^2\ln \left|X \right| ,\quad \\
C_2(\varphi )/c\sim 1-b_2 X^2+ b_3\left| X^3\right| ,\quad \\
C_4(\varphi )/c\sim 1-b_4 X^2+ b_5X^4+ b_6\left| X^5\right|.\quad
\end{array}
\end{equation}
Some of the coefficients $b_n$ will be specified later, for specific choices
of the PRM ensemble.

\subsection{Scaling and Level Derivatives.}

The statistics of BPs not only determines the statistics of Fourier
coefficients, but also those of the curvature $K(\varphi )=e^{\prime \prime
}(\varphi )$ and of higher-order derivatives as well. Several results are
known about the distribution of curvature in the case of stationary
ensembles \cite{grn,ZD93,vO94,HJS94}. Here we shall point out some immediate
conclusions that can be drawn from Fourier analysis, concerning the
statistics of an arbitrary level derivative $K_\alpha (\varphi )=e^{(\alpha
)}(\varphi )$ (being analytic, levels have derivatives of all orders), under
the scaling assumption (\ref{scal}).

{}From the very definition we get the mean square value of such a derivative
in the form:
\begin{equation}
\label{curv0}\overline{K_\alpha ^2}=\sum\limits_{-\infty }^{+\infty
}n^{2\alpha }|a_n|^2,
\end{equation}
where, as usual, the bar denotes an average over the phase $\varphi $. On
ensemble-averaging this equation we get

\begin{equation}
\langle\overline{K^2_{\alpha}}\rangle=\Delta^2\sum\limits_{-\infty}^{+%
\infty} n^{2(\alpha-1)}c_n
\end{equation}

{}From the scaling law we now obtain:
\begin{equation}
\label{Akk}\langle \overline{K_\alpha ^2}\rangle \approx \Delta
^2\sum\limits_{-\infty }^{+\infty }n^{2(\alpha -1)}{\sqrt{c}}P(n/{\sqrt{c}}%
)=A^{(2\alpha -2)}\Delta ^2c^\alpha
\end{equation}
where $A^{_{(2\alpha -2)}}$ is the $(2\alpha -2)$-th moment of the
distribution $P(t)$. From this equation, using the relation (\ref{mom})
between the
moments of $P(t)$ and the derivatives of $C(\varphi )$ at $\varphi =0$, we
obtain that the $(2\alpha -2)$-th derivative
of $C(\varphi )$ is
singular at $\varphi =0$ if, and only if,
the 2nd moment of the $\alpha $-th derivative has  nonintegrable
singularities as a function of $\varphi$.
On account of (\ref{singr}), this
occurs when $\alpha $ is larger than 1,2,3, respectively, for
GOE, GUE, GSE-like level repulsion. If, in addition, the PRM ensemble is
 stationary, then, for such values of $\alpha$, the  moments are divergent
at all $\varphi $.
For $\alpha =2$, that is, for ordinary curvature, and for  $\beta \geq 2$,
Eq.(\ref{Akk}) establishes the proportionality of rms curvature  to the
average kinetic energy:
$$
\langle\overline{K^2_2}\rangle^{1/2}\approx\sqrt{ A^{(2)}}\Delta c\approx
\pi\sqrt{2b_2}\Delta c
$$
where (\ref{mom}),(\ref{singr}) have been used to find $A^{(2)}$.
 This relation is similar to one found by Akkermans
and Montambaux \cite{AkkMont} for the case of a symmetry-breaking AB flux,
the difference being the additional $\varphi $-average performed here. If,
in addition, the ensemble is stationary, the same relation holds for the rms
curvature at any point $\varphi .$
\section{Stationary Ensembles}

\subsection{Spectral correlations}

The stationary ensembles we consider are defined by:

\begin{equation}
\label{pure}H(\varphi )=H_1\cos \varphi +H_2\sin \varphi,
\end{equation}
where $H_1,H_2$ are fixed, (i.e., $\varphi -$independent) random matrices,
drawn independently from one Canonical Gaussian ensemble (GOE, GUE, GSE).
Depending on the chosen ensemble, the PRM ensemble defined by (\ref{pure})
has trajectories in different matrix spaces (the space of real symmetric
matrices of rank $N$ for GOE; of complex Hermitean matrices of rank $N$ for
GUE; of symplectic matrices of rank $2N$ for GSE). For every $\varphi$, the
elements of the matrix $H(\varphi)$ are centered Gaussian random variables,
of variance

\begin{equation}
\label{norm}\left\langle \left| (H_{1,2})_{ij}\right| ^2\right\rangle =\frac
1{\beta N}\left( 1+\delta _{ij}\right) .
\end{equation}
With the chosen normalization of matrix elements, the average eigenvalue
density of $H(\varphi )$ obeys (in the $N\rightarrow \infty $ limit) the
Wigner semicircle law \cite{fritz} in the form
\begin{equation}
\rho (e)=\frac N{2\pi }\left( 4-e^2\right) ^{1/2}.
\end{equation}
In the central region of the spectrum the average level spacing is then $%
\Delta =\rho ^{-1}(0)=\pi /N$.

Because the equation (\ref{pure}) resembles the parametric equation of an
ellipse, the PRM ensembles just defined will be termed ''Gaussian Elliptic''
in the following. They have a number of distinguished features. In the first
place, they are stationary\cite{fritz,ZD93}; therefore, all ''pointwise''
distributions, such as the distribution of velocity $e^{\prime }(\varphi ),$
of curvature $e^{\prime \prime }(\varphi ),$ at given $\varphi ,$ are
independent of $\varphi $. Moreover, two-points statistical correlations
such as $\left\langle e^{\prime }(\varphi _1)e^{\prime }(\varphi
_2)\right\rangle $ only depend on the difference $\left| \varphi _2-\varphi
_1\right| .$

Secondly, at any $\varphi $ the derivative $H^{\prime }(\varphi )$ is
statistically independent of $H(\varphi )$, and belongs to the same
canonical ensemble; indeed, a simple computation shows that the matrix
elements of both $H(\varphi )$ and of $H^{\prime }(\varphi )$ are Gaussian
distributed, and that they are uncorrelated. As a consequence, the level
velocity $e^{\prime }(\varphi )$ has a Gaussian distribution. In fact, $%
e^{\prime }(\varphi )$ is given by an expectation value $\langle e|H^{\prime
}(\varphi )|e\rangle $ on the eigenvector $|e\rangle $ of $H(\varphi )$
corresponding to the eigenvalue $e(\varphi )$. On account of the
independence of $H(\varphi )$ and $H^{\prime }(\varphi )$, and of the
rotational invariance of the Gaussian ensembles, such a matrix element must
have the same distribution as any diagonal element of $H^{\prime }(\varphi )$%
. Therefore it has a Gaussian distribution, with variance

\begin{equation}
\label{varia} \left\langle e^{\prime}(\varphi )^2\right\rangle = \frac
{2}{\beta N}.
\end{equation}
An independent (albeit more complicated) proof of the Gaussian distribution
of level velocities can be obtained from supersymmetric calculations \cite
{Alts}. The variance (\ref{varia}) also follows from a statistical
mechanical analysis of level dynamics (Eq.(3.8) of \cite{ZD93}) under the
assumption of the applicability of the canonical ensemble.

Combining (\ref{varia}) with the mean spacing $\Delta $ we obtain for the
center of the spectrum the average variance of level velocity (also called
the average squared current)
\begin{equation}
\label{cn}c=\frac 2{\beta \pi ^2}N.
\end{equation}

We have numerically computed Fourier transforms of eigenvalues $e(\varphi )$%
, for matrices of the above type, of rank $N=40\div 400.$ The eigenvalues
were unfolded in order to normalize them to unit level spacing. In computing
(fast) Fourier transforms we have used grids of $128 \div 1024$ points in $%
(0,2 \pi)$ - the optimal size of a grid grows with the matrix rank $N$. To
avoid confusions, in the following the results of averages over $\varphi$
taken for a fixed level and a fixed realization will be called "mean
values", while by "average values" we shall mean results of statistical
averaging (over realizations, and/or over different levels). Thus "mean"
values will be fluctuating quantities, in contrast to "average" values. For
each level we have computed Fourier coefficients $a_n$, mean correlation
Fourier coefficients $r_n$ [defined by Eq.(\ref{rfi})], and the mean squared
velocity $r=\sum r_n$. Finally, statistical averages were taken
over all the eigenvalues lying in the central half of the spectrum, and the
results were further averaged over samples of $20 \div 300$ matrices.

As a first step, we have checked the scaling law (\ref{scal}). The
numerically computed moment $\langle\vert n\vert\rangle$, defined by (\ref
{scal2}), grows as $\sqrt N$, hence, as $\sqrt c$, as predicted by the
scaling law. Moreover, the relative width of the Fourier spectrum, measured
by $\delta_n:=(\langle n^2\rangle -\langle\vert n\vert\rangle^2)^{1/2}/
\langle\vert n\vert \rangle $ is constant, as required. Direct evidence for
the scaling law in the GUE case is presented in Fig.\ref{Fig.2}, where the
function $P_2(t)$ is plotted for various matrix sizes $N$ against the
rescaled variable $t=n/\langle\vert n\vert \rangle$.

All the assumptions which underlie the argument developed in the previous
Section appear to be satisfied for Elliptic ensembles, so that the small- $%
\varphi $ behaviour (\ref{singr}) or, equivalently, the large-$t$ behaviour (%
\ref{tail}) are expected.

Some of the coefficients $b$, which belong to the nonsingular part of the
expansion (\ref{singr}), are explicitly known:
\begin{equation}
\label{bb}b_2=2 ,\quad b_4=\frac{8}{3}
\end{equation}
as can be obtained \cite{jz} by expanding $C(\varphi)$ around $\varphi=0 $
to the lowest nonvanishing order, and using for the variance of curvatures, $%
\left\langle e^{\prime\prime }(\varphi )^2\right\rangle $, the values
derived from exact curvature distributions \cite{ZD93,vO94,HJS94}. The
coefficient $b_2$ has also been found via supersymmetric calculations \cite
{Alts}. In the case of rank-2 matrices we have exactly and explicitly
calculated the functions $C_1,C_2$, which are reported in the Appendix and
have the predicted small-$\varphi$ behaviour.

In numerically analyzing the large-$n$ behavior of the coefficients $c_n,$
two competing numerical effects, which obscure the real asymptotics, have to
be taken into account. The first is ''aliasing'', an artifact originated by
the discretization used in Fourier transforms, which tends to bend the
numerical tails upward. A second effect, which tends to bend them downward,
is due to finite statistics: however large the sample of matrices, the
avoided crossings sampled in it have a finite nonzero minimum width, that
will be detected by the Fourier transform as soon as $n$ becomes large
enough. To get rid of such effects, we have just restricted attention to
that part of the $c_{n\text{ }}$sequence that proved stable against
increasing the Fourier transform basis and the size of the sample. Results
are shown in Fig.\ref{Fig.3} for the three cases $\beta =1,2,4$, and agree
with the prediction $c_n\sim n^{-2-\beta }.$

Finally, we have used the numerical Fourier transforms to reconstruct the
scaling functions $\Phi _\beta (X)$. Results are shown in Fig.\ref{Fig.4} for
the three Elliptic ensembles, and in more detail in Fig.\ref{Fig.5} for GUE .
In this case we have compared the empirical $\Phi _2$ with the theoretical
predictions, eqs.(\ref{singr}) at small $X$, and $X^{-2}$ at large $%
X $, finding a good agreement in both cases.

\subsection{Statistics of Fourier coefficients}

An arbitrary rotation $\varphi \longmapsto \varphi +\alpha $ results in
multiplication of the Fourier coefficients $a_n$ by the phase factor $%
e^{in\alpha }$. On the other hand, such a rotation must leave the statistics
of the stationary ensemble unchanged. Hence $\langle a_n\rangle =e^{in\alpha
}\langle a_n\rangle $, and, since $\alpha $ is arbitrary, $\left\langle
a_n\right\rangle =0.$ A quite similar argument shows that $\left\langle
a_n^{*}a_m\right\rangle =0$ if $n\neq m$, which means that the complex
variables $a_n$ are pairwise uncorrelated. Nevertheless, they {\it cannot }%
be statistically independent. In fact, for any fixed realization they decay
exponentially fast at large $n$, but their variances decrease algebraically
instead. This sharp contrast between the individual and the average decay
entails a strong statistical dependence between coefficients $a_n$ at large $%
n$.

Formula (\ref{a}), shows that $a_n$ are sums of a large number $\sim \sqrt{c}%
\sim \sqrt{N}$ of contributions $I_j$, that come from the different BPs. On
assuming these contributions to be independent, one would naturally expect a
Gaussian distribution for $a_n$, which implies an exponential distribution $%
(\chi _2^2)$ for $|a_n|^2$. Numerical computations have confirmed this
prediction for all three universality classes and $n=1,2,3$ with $N=50$.
However, at very large $n$ the asymptotics (\ref{a},\ref{int2}) holds, with
only a few terms surviving in the sum, corresponding to very small, and
unlikely, avoided crossings. In this case a Gaussian distribution can hardly
be expected. A Gaussian distribution of {\it all} the Fourier coefficients
can be in fact ruled out: for otherwise these coefficients, being
uncorrelated, would also be independent, a case that we have just excluded.
In addition, other variables linearly depending on the $a_n$ such as, e.g.
the curvature $K(\varphi )=-\sum n^2a_ne^{in\varphi }$ should have a
Gaussian distribution%
\footnote{In fact, in the GUE, GSE cases the asymptotic behaviour of
$\langle\vert a_n\vert^2\rangle$ entails that the series for $K$
is mean-square convergent, so that a Gaussian distribution of all
terms in the series would enforce a Gaussian distribution of its sum.
}, which is known to be false\cite{ZD93,vO94,HJS94}.

Our numerical data have revealed broad, non Gaussian tails of the
distributions of the real and the imaginary parts of $a_n$ for sufficiently
large $n$. In Fig.\ref{Fig.6} we illustrate the transition from Gaussian to
non-Gaussian distribution of $a_n$, which occurs as $n$ increases, by
plotting the logarithmic variance (i.e., the variance of $\ln (|a_n^2|)$)
versus $t=n/\langle |n|\rangle ,$for the case of Elliptic GUE. The
logarithmic variance corresponding to a Gaussian distribution of $a_n$ is
equal to $\pi ^2/6$; this value is indeed found at small $n<\langle
|n|\rangle $ (for $N=40$, $\langle |n|\rangle \approx 12$), but not at
larger $n$, yielding evidence that the distribution of Fourier coefficients
is not Gaussian any more. Moreover, the same Fig.\ref{Fig.6} shows that the
log-variance does not scale with $n/\langle |n|\rangle $. Therefore,
although the 2nd moments of the Fourier coefficients scale in a simple way (%
\ref{scal}), uniformly (that is, for all $n$), the same is not true for
their global statistical distribution, except perhaps in the range $n<\sqrt{c%
}$, where, due to the Gaussian character of the distribution, the scaling of
2nd moments entails the scaling of all moments.

\subsection{Curvatures, and mean kinetic energy}

Eqn.(\ref{Akk}), and its consequences, are obviously valid in the case of
Gaussian Elliptic ensembles. In addition, $\langle\overline{K^2_{\alpha}}
\rangle=\langle K^2_{\alpha}(\varphi)\rangle$ for all $\varphi$ because of
the stationarity of the ensemble. From Eqs. (\ref{singr})(\ref{mom})(\ref{bb})
one finds $A_2^{(2)}=4\pi^2$ and $A_4^{(2)}=16 \pi^2/3$. Thus, in
particular, in the GUE Elliptic ensemble the root mean square (rms) of the
curvature is, at any point $\varphi$, proportional to $c$, with a factor $2
\pi \Delta $.

The Gaussian distribution of the level velocities entails a Porter--Thomas
distribution of squared velocities. Nevertheless, such a distribution cannot
be expected for the mean square level velocity $r=\sum r_n$ (the square
level velocity, averaged over $\varphi $; a realization- dependent quantity,
see Eq.(\ref{RR})), which is {\it not} the square of a Gaussian variable.
Braun and Montambaux \cite{BrMont} have numerically studied the distribution
of $r$ for the 3D Anderson model in the metallic regime, and have
approximated their data by a log-normal distribution. In this Section we
describe our results for the distribution of $r$ in the Elliptic GUE
ensemble.

Our data show that the variance $\sigma^2_r$ of $r$ is proportional to $%
N^{3/2}$, while the logarithmic variance $\sigma^2_{r\ln}= \langle
\ln^2(r)\rangle-\langle \ln(r)\rangle^2 $ scales as $N^{-1/2}$. Thus the
relative fluctuation of $r$ vanishes in the limit $c\to\infty$, that is, $%
r/c $ is a self-averaging quantity. Figure \ref{Fig.7} displays the
distribution $P(\ln(r))$ of the logarithm of $r$ obtained from $200$ GUE
matrices with $N=50$. The small but significant asymmetry with respect to
the mean advises against a Gaussian fit; as a matter of fact, a much better
fit for the empirical distribution of $r$ was obtained by means of a $%
\chi^2_{\nu}(r)$ distribution, with the average $\langle r\rangle=c$, and
with the number of degrees of freedoms $\nu\approx 2/ \sigma^2_{r\ln}$. The
latter estimate follows from the observation that the logarithmic variance $%
\sigma^2_{\ln}$ of the $\chi^2_{\nu}$ distribution does not depend on the
mean and is given by $\Psi^{\prime}(\nu/2)$ where the Trigamma function \cite
{atlas} $\Psi^{\prime}(\nu/2)\sim 2/\nu$ at large $\nu$. The resulting
distribution (solid line in the figure) fits the data much better than the
log-normal distribution. A similar distribution describes the mean kinetic
energy statistics for GOE and GSE.

As already mentioned, $\sigma _{r\ln }^2$ depends on the matrix size as $%
N^{-1/2}$. Hence $\nu $ is proportional to $\sqrt{N}$ and also to the mean
coefficient $\langle n\rangle $; in other words, the number of degrees of
freedom is proportional to the width of the Fourier spectrum. Numerical data
give for GUE $\nu \approx 3\langle |n|\rangle $. Therefore, our results
appear to indicate that the statistics of $r=\sum n^2|a_n|^2$ are determined
only by the Fourier amplitudes $a_n$ with $n<\sqrt{c}$, which, as we know,
are Gaussian and uncorrelated. This would also explain the observed
dependence of the variance of $r$, for
\begin{equation}
\label{sig2}\sigma _r^2\sim \sum\limits_{|n|<{\sqrt{c}}}c_n^2\sim
c^{3/2}\int\limits_0^1P^2(t)dt
\end{equation}

This interpretation is far from complete, though. In fact the scaling law
implies that the relative contribution to $r$ of the region $n>\sqrt{c}$ is
independent of $c$, so that there is no ''a priori'' reason why it should be
negligible.

\section{Periodic Band Random Matrices}

\subsection{General Properties}

Let us consider a circular array of $N$ sites on a ring, labelled by the
index $j=1,...N.$ By definition, a Periodic Band Random Matrix has nonzero
matrix elements $H_{jk}(\varphi )$ only for sites which are no more than $b$
sites apart {\it on the circle, }where $b\leq N$ is an integer specifying
the width of the band. The matrix elements are further specified by:
$$
\begin{array}{c}
H_{jk}(\varphi )=h_{jk}
\text{ if }\left| j-k\right| \leq b \\ H_{jk}(\varphi )=h_{jk}e^{i\varphi }
\text{ if }j\leq k+b-N \\ H_{jk}(\varphi )=h_{jk}e^{-i\varphi }\text{ if }%
j\geq k+N-b,
\end{array}
\label{pbrm}
$$
where $h_{jk}$ are real Gaussian variables, independent of $\varphi $, $%
h_{jk}=h_{kj}$, with the normalization

\begin{equation}
\label{pnorm}\big\langle h_{jk}^2\big\rangle =\frac{N+1}{2bN}\bigl( 1+\delta
_{jk}\bigr).
\end{equation}
For an odd value of $N$ and $b=(N+1)/2$ we get at $\varphi =0$ a full GOE
matrix.

PBRMs are obtained from a Bloch decomposition of infinite, periodic, banded
random matrices, and $\varphi $ is the corresponding Bloch index. The PBRM
ensemble differs from the periodic random matrix ensembles matrices
considered above in two important respects:

\begin{enumerate}
\item  {\ $H(\varphi )$ exhibits a proper diffusive regime in the range $%
1<<b<<N<<b^2$ (which should not be identified with the ballistic regime $%
b\approx N$), and a localized regime at $N>b^2.$ All statistical properties
of $H(\varphi )$ depend on $b$ and $N$ through the scaling variable $%
x=b^2/N, $ which is proportional to the ratio between the localization
length of the eigenfunctions in the limit $N=\infty $ and the ''sample
size'' $N$\cite{PBRM}; }

\item  {\ The self-adjoint matrix $H(\varphi )$ satisfies $H(-\varphi
)=H^{*}(\varphi )$. Therefore at $\varphi =0$ and $\varphi $$=\pi $ the
matrix $H(\varphi )$ is real symmetric, while at any other value of $\varphi
$ it is complex-Hermitean. This is exactly the behaviour expected for an
Hamiltonian depending on an Aharonov-Bohm magnetic flux.}
\end{enumerate}

On account of both the properties mentioned above, PBRMs provide a model for
the dynamics of electrons in small disordered rings threaded by a magnetic
flux $\varphi $ (in dimensionless units). Previous studies \cite{PBRM} have
shown that the behaviour of the ensemble-averaged, zero-flux scaled absolute
curvature ${\cal K}:=\Delta ^{-1}\left\langle \left| e^{\prime \prime
}(0)\right| \right\rangle $ is quite similar to the one expected of the
average conductance in the scaling theory of localization. The distribution
of ${\cal K}$ also depends on $x$, being identical to the generalized Cauchy
distribution (first conjectured by Zakrzewski and Delande\cite{ZD93}, and then
proven for GUE \cite{vO94} and for a broad class of ensembles of random
matrices \cite{HJS94}) in the delocalized limit $x\to \infty $, and
resembling a log-normal distribution in the opposite localized limit. In
general, all statistical properties change continuously as the localization
parameter varies from the localized to the delocalized limit. For instance,
Fig.\ref{Fig.8} shows the $x-$ dependence of the average squared current $c$%
, the variance $\sigma _r^2$, the log-variance $\sigma _{r\ln }^2$, and the
moment $\langle |n|\rangle $ of the Fourier spectrum. The latter is close to
unity in the localized regime; but in the diffusive regime it scales as $%
\sim x^{1/2}$. This result has a simple explanation: the Thouless
conductance $\left\langle \left| K(0)\right| \right\rangle \Delta ^{-1}$
obeys Ohm's law in the diffusive regime \cite{PBRM}, hence it is
proportional to $x$. On the other hand, according to a relation discussed by
Akkermans and Montambaux \cite{AkkMont} and numerically verified for PBRM%
\cite{PBRM} and 3D Anderson model\cite{anders}, in that regime the Thouless
conductance is also proportional to $C_{AB}(0)$. Therefore, $c\sim x$, and $%
\langle n\rangle \sim \sqrt{c}\sim \sqrt{x}$. For the same reason, on
approaching the extreme delocalized regime some power-law dependencies on $%
c\sim x$ are recovered, which are typical to canonical ensembles of ''full''
random periodic matrices: e.g., $\sigma _r^2\sim x^{3/2}$, $\sigma _{r\ln
}^2\sim x^{-1/2}$ and $\delta _n=const$.

Though the general approach described in Section II is still valid, some
modifications of the picture described for the Elliptic ensembles are
imposed by the specific features of the PBRM ensemble; in the first place,
by the symmetry breaking occurring at $\varphi=0$, due to which this
ensemble is not stationary. One-point spectral statistics (level spacing
distributions, curvature distributions, ...) change as $\varphi $ is
increased from $0$ to some value $\varphi _c$ which is however small, in the
sense that it decreases with $1/\sqrt c$; at large $c$ the transition is
therefore a very sharp one\cite{PM}. The correlation analysis of Sec. II
cannot be literally replicated for the PBRM model, as the distribution of
BPs in the complex $\varphi -$plane cannot be homogeneous in $\varphi;$ the
width $\gamma $ of a complex branch-point may not be statistically
independent on its location $\varphi _0,$ because of the changes of symmetry
occurring at special values of $\varphi .$ We have, therefore, to
investigate how the statistical distribution of BPs is affected by the
symmetry breaking. This is in turn closely related to the distribution of
avoided crossing sizes (gaps) and their positions, that will be analyzed
below.

\subsection{Avoided Crossings.}

The above mentioned symmetry of the PBRM model entails that eigenvalues $%
e(\varphi )$ are symmetric with respect to $\varphi =0,\pi $ ($\varphi =\pi $%
),  and their velocities vanish there. It follows that relative extrema for
the level spacings are observed at these symmetry-breaking values of $\varphi
$%
. In other words, a large set of avoided crossings (relative {\it minima} of
spacings) must appear at a {\it fixed} value of $\varphi =0$ (or $\pi $).
Thus, with probability one, there will be BPs lying on the lines $\varphi
=0,\pi $. Such BPs will contribute a delta-function term in the distribution
of real parts $\varphi _0$ of BPs. Moreover, since the width of any of these
avoided crossings is at once a nearest-level spacing, the distribution of
these widths, at small values, will behave exactly like the level spacing
distribution at $\varphi =0,\pi $, hence with the exponent $\beta =1$ (and
not $\beta -1$, as with BPs having a random position).

We therefore infer that the symmetry breaking lines contribute a singular
term to the distribution $p(\gamma ,\varphi )$ of BPs in the complex plane,
of the form (apart from a normalization constant):
\begin{equation}
\label{delta}p_s(\gamma ,\varphi )=(\delta (\varphi )+\delta (\varphi -\pi
))f_0(\gamma )
\end{equation}
with $f_0(\gamma )\sim \gamma $ at small $\gamma $. We shall now examine the
other type of avoided crossings, namely those which have fluctuating positions
{\it inside} the $%
(0,\pi )$ interval.
In order to numerically compute the positions and sizes of avoided crossings
we have used the same procedure as in previous studies on stationary
ensembles; we refer the interested reader to \cite{ZDK93} for details.

In fig.\ref{Fig.9} we illustrate the distribution of the positions of the
"fluctuating"
avoided crossings lying inside $\left( 0,\pi \right) $, for matrices
belonging to the diffusive regime. A strong dependence of the abundance of
avoided crossings with respect to $\varphi $ is apparent. In particular,
 fluctuating BPs
appear to be repelled by the symmetry breaking values
(where, as remarked above, other BPs already exist.)
. The abundance of fluctuating avoided crossings depends on the position
$\varphi$ in a way which resembles that of the average squared level
velocity, which will be discussed later.

Although the distribution of avoided crossings is not homogeneous in $%
\varphi $, the (conditional) distribution of their sizes appears to be the
same at all $\varphi\not= 0,\pi$ and is presented as an insert in fig.\ref
{Fig.9}. Thus we argue that the BPs located inside $(0,\pi)$ contribute a
"smooth" part $p_0(\gamma,\varphi)$ to the total density of BPs, which is
not $\varphi$-independent in the vicinity of $0,\pi$, but is $\sim \gamma$
at small $\gamma$, at all $\varphi$.

\subsection{Spectral Correlations}

On the grounds of the above discussion, the distribution of the BPs in the
complex plane consists of a ''smooth'' part, given by the BPs internal to $%
(0,\pi )$, plus a ''singular'' part, exactly located on the symmetry
breaking lines. {\it In both parts, the probability of small $\gamma $'s
scales with the same exponent $1$}. The correlation analysis of Section II
now applies unaltered to the effect that the type of singularity exhibited
by $C_{AB}(\varphi )$ at $\varphi =0$ is of GUE type, like that of $%
C_2(\varphi )$; the numerical coefficients, nevertheless, may be different
(and will indeed become different on moving towards the localized regime).

On numerically computing Fourier transforms, we have in fact found
distributions $c_n$ whose general features cannot be distinguished from
those observed in the pure GUE case, on the level of accuracy of our
computations. In particular, a decay $c_n\sim n^{-4}$ is apparent (figs.\ref
{Fig.10},\ref{Fig.12}). The corresponding correlation functions are
almost identical to those obtained for the Elliptic GUE case
(fig.(\ref{Fig.11}).

The $n^{-4}$  decay is  by no means consistent with the
behaviour suggested by Braun and\ Montambaux\cite{BrMont}, which is instead
of the type exhibited by $C_1(\varphi )$ in (\ref{singr}) and  would
therefore result in a $n^{-3}$ decay.
Berry and Keating (BK) \cite{BK94} have proposed an explicit semiclassical
approximation for $C_{AB}$, which is compared with the numerically computed
CF for the PBRM model in fig.\ref{Fig.10}. A free parameter, $w^{*}$, which
appears in the BK formula, has been fixed by means of a one-parameter fit to
our data for $N=201$, $b=64$. The obtained value $w^{*}=3.92$ is quite
small; therefore, we have used Eq.(28) of ref.\cite{BK94}, instead of its
limiting form, Eq.(29), which is valid in the limit of large $w^{*}$.
Including the ''non-diagonal'' correction \cite{BK94} yields no significant
improvement to the fit. The minimum of the semiclassical CF is too shallow
as compared to numerical data (see  inset in fig.\ref{Fig.10}). Moreover,
since the $C_{AB}$ of BK is an analytic function at all real $\varphi $, its
Fourier transform has an exponentially decaying tail, at variance with the
real algebraic tail (main panel in fig.\ref{Fig.10}).
\footnote{The value of $w^*$ can also be estimated from the large $\varphi$
behaviour, Eq.(39) of ref.\cite{BK94}; a much larger value is found in this
way. Though the  large $\varphi$
behaviour is now correct, the small and intermediate $\varphi$ behaviour
is in even
stronger disagreement with numerical data.}

It is worth pointing out, as a purely empirical observation, that a
significant {\it de facto} agreement with numerical data is provided by a
recently proposed semi-empirical representation of the CF \cite{jz} , which
we show in the same inset. No fitting was used in this case, the only
adjustement being the rescaling from the dimensionless parameter $X$ (for
which the semiempirical representation of CF was given) to $\varphi .$
Different data sets (e.g. for $N=201$, $b=101$ , close to the ballistic
regime, with $w^{*}\approx 6.35$) have shown an even better agreement with
this analytical representation (while no significant improvement has been
observed with the BK formula).

Nevertheless this semi-empirical representation , too, is in the form of an
analytic function, and cannot therefore correctly represent the small- $%
\varphi $ behaviour. Like the BK formula, in the Fourier domain it yields
exponential instead of algebraic tails, at variance with numerical data.

On moving towards the localized regime, the $C(\varphi )$ curve undergoes a
continuous deformation (fig.\ref{Fig.12}), and the singularity at $\varphi
=0 $ remains of the same type all the way down to the deeply localized
regime; in fact, Fourier transforms always exhibit a $|n|^{-4}$ tail. In all
the investigated range we have obtained evidence that $C(\varphi )$ depends
on $b,N$ only through the localization ratio $x=b^2/N$: $C(\varphi
,b,N)\approx C(\varphi ,x)$ at $b,N>>1$. In the diffusive regime, $x>1$, the
scaling (\ref{scal}) is still approximately valid: $C(\varphi ,x)\approx
c(x)\Phi _{AB}(\pi \varphi {\sqrt{c(x)}})$ with $c(x)=C(0,x)\sim x$ (weak-
localization corrections are hardly assessed at the present level of
accuracy). As an illustration, we plot in fig.({\ref{Fig.13}}) the position $%
\varphi _0(x)$ of the zero of $C(\varphi ,x)$, as a function of $x$. In the
diffusive part of the plot, $\varphi _0(x)$ is proportional to $\sqrt{x}$,
consistently with (\ref{scal}) and with $c\sim x$. In the localized regime
this scaling is not valid any more. The Fourier spectrum shrinks
exponentially fast; all the harmonics except the 1st tend to disappear, so
that $C(\varphi )$ tends to $c\cos \varphi $.

\subsection{Statistics of Fourier coefficients}

In the PBRM ensemble, the Fourier coefficients $a_n$ are real. Our numerical
data for their statistical distributions show in the delocalized regime a
behaviour similar to the one observed for stationary ensembles. At small $%
n<\langle n\rangle $ we obtained a Gaussian distribution of $a_n;$ as a
consequence, the log - variance $\sigma _{\ln }^2$ (i.e., the variance of $%
\ln $ $(a_n^2)$) is constant and equal to $\pi ^2/2.$ . For $n$ larger than $%
\langle n\rangle $ the distribution exhibits broad, exponential tails (fig.(%
\ref{Fig.14})).

On the other hand, in the localized regime, none of the coefficients is
Gaussian-distributed.

In the stationary case, the assumption that the phases $\varphi _{0j}$ of
the BPs come at random immediately leads to the vanishing of correlations
between different $a_n$ - a result which we have obtained in a different
way, and which is intimately linked to the stationarity of the ensemble. In
the PBRM case, which is not stationary, different Fourier coefficients have
nonzero correlations; the asymptotic formula (\ref{int2}) can be used
together with formula (\ref{a}) to study the correlation matrix $\langle
a_na_m\rangle $.

The random phase assumption is not legitimate for the BPs which have
nonrandom phases $0,\pi $. From (\ref{a}),(\ref{int2}) it follows that these
BP contribute a correlation

\begin{equation}
\label{anm}\langle a_na_m\rangle \sim (1+(-1)^{n+m})\langle \gamma {\frac{%
e^{-\gamma (|n|+|m|)}}{|nm|^{\frac 32}}}\rangle \sim (1+(-1)^{n+m})M{\frac
1{|nm|^{\frac 32}(|n|+|m|)^3}}
\end{equation}
with $M$ a numerical factor. In estimating the ensemble average, the
GOE-like distribution of small $\gamma $ of BPs lying on the symmetry lines
was used. We shall  neglect the contributions of the remaining BPs, which
have fluctuating phases. The formula (\ref{anm}) will be used in the next
Section to analyze the behaviour of certain quantities near the
symmetry-breaking point $\varphi =0$.

\subsection{Kinetic Energies and Curvatures.}

{}From the expansion
$$
\langle e^{\prime }{}^2(\varphi )\rangle =
\sum\limits_n\sum\limits_mnm\langle a_na_m\rangle e^{i(n-m)\varphi }
$$
we obtain, using (\ref{anm}),
$$
\langle e^{\prime }{}^2(\varphi )\rangle \ =\sum\limits_kB_ke^{2ik\varphi }
$$
where the coefficients $B_k$ have the asymptotic behaviour:
$$
B_k\sim M\sum\limits_p{\frac{|k^2-p^2|^{-\frac 12}}{max(|k|,|p|)^3}}\sim {%
\frac 1{|k|^3}}
$$
As we know, such an asymptotic behaviour of the Fourier expansion entails a
singularity at $\varphi =0$, of type $\langle e^{\prime }{}^2(\varphi
)\rangle \sim \varphi ^2\ln |\varphi |$ (and a similar behaviour close to $%
\varphi =\pi $), as predicted in \cite{KB94} for a GOE-GUE transition and
numerically observed for the Anderson model \cite{BrMont} \cite{PM}.

A quite similar analysis shows that the 2nd moment of the curvature has a
logarithmic singularity at $\varphi =0,\pi $, as again it was predicted in
\cite{KB94,BrMont}%
\footnote{We note in passing that this singularity
is an integrable one, and that this fact alone is sufficient to
exclude that the 2nd derivative of $C(\varphi)$ may be singular at
$\varphi=0$: see the remarks following eqn.\ref{Akk}.}

Figure \ref{Fig.15} shows the dependence of the average squared velocity
(normalized to its maximal value ) as a function of the scaling variable, $%
X=\pi \varphi \sqrt{c}$, for $N=71$ and for several band widths $b$. Our
numerical data are consistent with the mentioned logarithmic behaviour.
Moreover, in the delocalized regime, the data for different values of $b$
follow the universal dependence recently predicted for the GOE-GUE
transition \cite{TA94}

\begin{equation}
\label{tanig}\frac{\langle e^{\prime }(\varphi )^2\rangle }{e^{\prime
}{}_{max}^2}=1-\frac{\sqrt{\pi /2}{\rm erfi}(\sqrt{2}X)\bigl[1+2X^2\exp
(2X^2){\rm Ei}(-2X^2)\bigr]}{2X\exp (2X^2)},
\end{equation}
where Ei(.) denotes the exponential integral and erfi($x$)=-i erf(i$x$) is
the imaginary error function \cite{atlas}.

The statistics of the mean kinetic energy is similar to the one discussed
for Elliptic ensembles; again, the distribution $P(r)$ is better
approximated by a $\chi _\nu ^2$ distribution than by a log-normal one. In
the localized regime, the width of the distribution $P[\ln (r)]$ increases
with decreasing $x$. As shown in fig.\ref{Fig.8}, on decreasing $x$ the
logarithmic variance $\sigma _{r\ln }^2$ grows faster than $x^{-1/2}$.

The relation (\ref{Akk}) is still valid, as is the proportionality between
the width of the curvature distribution $P(K)$ at $\varphi =0$,
measured by $g_1=\langle |K(0)|\rangle $, and $c$ \cite{AkkMont}\cite{PBRM}.
The proportionality constant is close to $2\pi ,$ as  numerically observed for
the Anderson model \cite{BrMont} and  analytically found in \cite{HJS94}.
Localization effects produce deviations from both proportionality relations
at $x<<1$ \cite{anders}\cite{PBRM}.

Within the above described Fourier framework, it is also possible to analyze
how the distribution of the curvature $K(\varphi )$ depends on $\varphi $.
This problem is also attracting interest; here we shall just make some
qualitative remarks, deferring a more detailed analysis to future
publications. The curvature distribution is affected by the distribution of
BPs in the vicinity of the chosen point $\varphi $ on the real axis; in fact
BPs coming very close to this point determine large curvatures, hence their
statistical distribution determines the tail of the curvature distribution.
However, the singular and the smooth parts of the BP distribution contribute
in a different way in this tail. The singular part yields a relatively large
probability for BPs coming close to $\varphi =0,\pi $, which in turn yields
a large probability of large curvatures there; as a matter of fact, the
curvature distribution at $\varphi =0,\pi $ was found to be of the
Zakrzewski-Delande type \cite{PBRM}. Nevertheless the same BPs, being bound
to the lines $\varphi =0,\pi $, cannot come arbitrarily close to a point $%
\varphi \not =0,\pi $, and cannot therefore produce large curvatures there.
In fact, if {\it only} the singular part of the BP distribution is taken
into account, then a simple estimate based on (\ref{a}),(\ref{int2}),
(\ref{anm}) shows that
at $\varphi \not =0,\pi $ {\it all} the moments of $K(\varphi )$ are finite.
Therefore, in any ensemble for which a smooth part is missing, the tail of
the curvature distribution at $\varphi \neq 0,\pi $ decreases faster than
any power of curvature. This is exactly the case with the $2\times 2$ matrix
ensemble considered in \cite{KB94}, where {\it all} BPs lie at $\varphi =0$ (%
$\gamma =0$ in the notation of \cite{KB94}), and the curvature distribution
was in fact shown to have Gaussian tails at $\varphi \not =0.$

Short of being typical of ensembles exhibiting a magnetic flux-induced
GOE-GUE transition,such a behaviour rather appears as an artifact of
ensembles lacking the smooth part of the BP distribution; the presence of
such a part would in fact restore a chance for BPs coming close to {\it any
}point on the real line. The impact of this fact on the curvature
distribution is currently under investigation.

\section{Conclusions.}

In this paper we have illustrated the usefulness of analytic continuation to
complex parameter values in the analysis of various statistics of parametric
level dynamics, which are essentially determined by the statistical
distribution of complex level crossings, {\it i.e}, of the singularities of
branch-point type exhibited by the eigenvalues as functions of the complex
parameter.
Though this approach is not, for the time being, as powerful as other
methods (such as, e.g., supersymmetric ones), it offers, in our opinion, a
particularly perspicuous picture.

Our analysis was in many respects a heuristic one, although one rigorous
result was also given: independently of symmetry, the level velocity
correlation function can be analytic only if the matrix ensemble is so
built, as to make too narrow avoided crossings impossible. Thus in most
cases this correlation function has singularities, which have the same
origin as the divergence of the 2nd moments of certain level derivatives, to
which they are in fact connected by the relation (\ref{Akk}). The latter is
in a (rather superficial) sense a generalization of the Akkermans-Montambaux
relation for conductance.

Our estimates for the singularities of the correlation function were
confirmed by numerical simulations, and also cover the physically important
case, in which the parameter has the meaning of an Aharonov-Bohm flux. This
case was modelled by the Periodic Band Random Matrix ensemble. Unlike most
of the usually considered ensembles, which consist of ''full'' random
matrices, this ensemble exhibits a proper diffusive regime in a suitable
parameter range, marked by an ohmic dependence of the average curvature on
the matrix size; therefore the corresponding eigenfunctions, although
delocalized, have to differ from those of ''full'' random matrices in some
as yet not fully understood, but nonetheless essential respect. The
investigation of this ensemble has confirmed that various scaling
properties, originally established for conventional ensembles, remain
substantially valid in the diffusive regime (already investigated in \cite
{BrMont} for a different model).

The GOE-GUE breaking of symmetry which occurs on switching on the
Aharonov-Bohm flux appears to have no impact on the behaviour of the
correlation
function, which looks the same as for a homogeneous GUE ensemble.
Nevertheless, the transition is reflected by a singular behaviour of other
quantities. Some features of this transition, such as e.g. the way it
affects the curvature distribution, are not yet understood, as they may not
be correctly reproduced by $2\times 2$ matrix ensembles, which have a
nongeneric distribution of branch points.

For the case of stationary ensembles, we have shown numerically
that, whereas the Fourier amplitudes $a_n$ of the eigenvalues
 are approximately Gaussian distributed when $n$ is small in comparison to the
width of the Fourier spectrum,
 strong deviations from Gaussian distribution occur at large $n$.
This behaviour can be qualitatively connected with the statistical distribution
of level derivatives, which have Fourier expansions in which the role of the
non-Gaussian
amplitudes $a_n$ is enhanced, the more, the higher the order of the derivative.
While 1st order derivatives (level velocities) are still
Gaussian distributed, this is no longer true already for 2nd order derivatives
(level curvatures), which have in fact a Cauchy distribution; a strongly
non-Gaussian distribution is therefore to be expected for all higher-order
derivatives.

Our theoretical apparatus heavily hinges on periodicity, and it is a natural
question, whether any of our conclusions can be generalized to the general
case of a non-periodic parameter dependence. Some kind of a stationary
behaviour is needed, in order that a velocity correlation may be defined; an
essential step is then to filter out any "secular" component of the
eigenvalue motion, as can be done, e.g., by the unfolding process. In any
case, the resulting branch-point pattern and the corresponding Fourier
analysis may be, in our opinion, significantly different from the periodic
case discussed in this paper.

We thank Felix Izrailev for bringing the problem of spectral correlations
in banded random matrices to our attention.
I.G. acknowledges an influential discussion with
Eric Akkermans.
K.{\.Z}. enjoyed fruitful discussions with
Yan Fyodorov, Marek Ku\'s and Dima Shepelyansky, and is
 thankful to all colleagues of the University of
Milano, sede di Como, for their hospitality during his stay in Como.
Financial support by the project nr 2-P30210206 of
Polski Komitet Bada{\'n} Naukowych is gratefully acknowledged.
Laboratoire Kastler-Brossel, de l'Ecole
Normale Sup\'erieure et de l'Universit\'e Pierre et Marie Curie, is
Unit\'e Associ\'ee 18 du Centre National de la Recherche Scientifique.

\section{ Appendix}

We compute analytically the velocity correlation for the level dynamics of $%
2\times 2$ matrices $H(\varphi )=H_1\cos\varphi +H_2\sin \varphi $, where
$$
H_1=\left(
\matrix { a_1 & {1\over \sqrt 2}(x_2+ix_3) \cr
	      {1\over \sqrt 2}(x_2-ix_3) & a_2\cr }\right) ,\quad H_2=\left(
\matrix { b_1 & {1\over \sqrt 2}(y_2+iy_3) \cr
	   {1\over \sqrt 2}(y_2-iy_3) & b_2\cr }\right) \eqno {(A.1)}
$$
both belonging to GUE or GOE ensembles; in the latter case $x_3=y_3=0$. All
matrix elements are normally distributed with the same variance. Setting $%
a_1+a_2=\sqrt{2}a$, $b_1+b_2=\sqrt{2}b$, $a_1-a_2=\sqrt{2}x_1$ and $b_1-b_2=
\sqrt{2}y_1$, we have the following expressions for the eigenvalues:
$$
e_{\pm }(\varphi )={\frac 1{\sqrt{2}}}(a\cos \varphi +b\sin \varphi )\pm {%
\frac 1{\sqrt{2}}}|\vec x\cos \varphi -\vec y\sin \varphi |\eqno {(A.2)}
$$
We begin by computing the following eigenvalue correlator, which actually
depends on the difference $\varphi =\varphi _2-\varphi _1$:
$$
F(\varphi )=\left( {\frac \omega \pi }\right) ^{n+1}\int
da\,db\,d^nx\,d^nye^{-\omega (a^2+b^2+|x|^2+|y|^2)}{\frac 12}[e_{+}(\varphi
_1)e_{+}(\varphi _2)+e_{-}(\varphi _1)e_{-}(\varphi _2)]\eqno {(A.3)}
$$
with $n=2,3$ respectively for GOE and GUE. The velocity correlator is then
obtained as a second derivative: $C(\varphi )=-F"(\varphi )\Delta^{-2}$.

The eigenvalue correlator is the sum of two contributions $F_1 + F_2$, where
$$
F_1 = {\frac{\omega}{\pi}}\int da \, db e^{-\omega (a^2+b^2)} {\frac{1}{2}}
(a\cos\varphi_1 + b\sin\varphi_1)(a\cos\varphi_2 + b\sin\varphi_2)= {\frac{1
}{{4\omega}}}\cos\varphi%
$$
$$
F_2 = \left ({\frac{\omega}{\pi}}\right )^n \int d^nx \, d^ny e^{-\omega
(|x|^2+|y|^2)}{\frac{1}{2}} |\vec x \cos \varphi_1-\vec y \sin \varphi_1
|\cdot |\vec x \cos \varphi_2 - \vec y \sin \varphi_2 |
$$
With the linear change of variables, for $\varphi $ different from $0$ and $%
\pm\pi$,
$$
\vec u= {\frac{{\sqrt\omega}}{{\sin\varphi}}} (\vec x \cos \varphi_1-\vec y
\sin \varphi_1 )\quad,\quad \vec v = {\frac{{\sqrt\omega}}{{\sin\varphi }}
}
(\vec x \cos \varphi_2 - \vec y \sin \varphi_2 )
$$
the second integral simplifies to
$$
F_2= {\frac{1}{{2\pi^n\omega}}} |\sin\varphi |^{n+2} \int d^n u |u|\int d^n
v |v|e^{-|u|^2-|v|^2 +2\vec u\cdot\vec v \cos\varphi} \eqno {(A.4)}
$$
Note that $F_2$ is an even function of $\varphi $.

Let us first consider the case of GOE. In polar coordinates, letting $\theta
$ be the angle between the two vectors $\vec u $ and $\vec v$, we have
\begin{equation}
\begin{array}{c}
F_2(\varphi )=
{\frac{1}{{2\pi^2\omega }}} \sin\varphi^4 2\pi \int_0^\infty u^2 du
\int_0^\infty v^2dv e^{-u^2-v^2}\int_0^{2\pi} d\theta
e^{2uv\cos\varphi\cos\theta } \\ = {\frac{2}{\omega }}\sin ^4\varphi
\int_0^\infty u^2 du \int_0^\infty v^2dv e^{-u^2-v^2} I_0(2uv\cos\varphi )
\end{array}
\end{equation}
where $I_0(x)$ is a Bessel function. The double integral is tabulated
[Prudnikov, vol. 2]and we obtain:
$$
F_2(\varphi )={\frac{1}{{4\omega }}}
\left [ 2E(\cos\varphi )-\sin^2\varphi K(\cos\varphi) \right ] $$
where $E$ and $K$ are the complete elliptic functions. Adding the $F_1$ term,
and computing the second derivative we get the velocity correlator for GOE.
Explicitly, it is a function of $t=\cos\varphi$:
$$
{\frac{{C(\varphi)}}{{C(0)}}} = {\frac{1}{{2t^2}}}
\left [ t^3 +E(t)-(1-t^4)K(t) \right ]\eqno {(A.5)}
$$
The small $\varphi $ expansion has the expected logarithmic term:
$$
{\frac{{C(\varphi)}}{{C(0)}}} = 1 -\varphi^2 \left ( {\frac{3}{2}}\log 2 -
{\frac{%
1 }{8}} - {\frac{3}{4}}\log |\varphi | \right ) + {\cal O}(\varphi ^4 ) \eqno
{%
(A.6)}%
$$
At $\varphi =\pi/2$ $(t=0)$, the normalized correlator has the value $-\pi/8$,
and at $\varphi=\pi-\epsilon $ it vanishes as $-\epsilon ^2 [(1/8) +(3/4)\log
(4/ \epsilon )]$.

Let us now consider the GUE case. In spherical coordinates, letting $\theta $
be the angle between the vectors $\vec u$ and $\vec v$, we have:
\begin{equation}
\begin{array}{c}
F_2(\varphi )=
{\frac 1{{2\pi ^3\omega }}}|\sin \varphi |^58\pi ^2\int_0^\infty
u^3du\int_0^\infty v^3dve^{-u^2-v^2}\int_0^\pi d\theta \sin \theta
e^{2uv\cos \varphi \cos \theta } \\ ={\frac 1{{\pi \omega }}}{\frac{{|\sin
\varphi
|^5}}{{\cos \varphi }}}\int_0^\infty du\int_0^\infty dv\sqrt{uv}e^{-u-v}{\rm %
sinh}(2\sqrt{uv}\cos \varphi )
\end{array}
\end{equation}
The double integral can be related to the following, available in the tables
[Prudnikov, vol 1 pag 573] $(pq>1/4)$:
$$
\int_0^\infty dx\int_0^\infty dy{\rm cosh}(\sqrt{xy})e^{-px-qy}={\frac 4{{%
4pq-1}}}+{\frac 4{{(4pq-1)^{3/2}}}}{\rm arcsin}\left( {\frac 1{{2\sqrt{pq}}}}%
\right)
$$
For $0<\varphi <\pi $ the result is
$$
F_2(\varphi )={\frac 1{{2\pi \omega }}}\left[ \left( {\frac \pi 2}-\varphi
\right) \left( 3\cos \varphi +{\frac{{\sin {}^2\varphi }}{{\cos \varphi }}}%
\right) +3\sin \varphi \right]
$$
After summation of the $F_1$ term, and derivating twice we obtain the
velocity correlation for $2\times 2$ GUE matrices:
$$
{\frac{{C(\varphi )}}{{C(0)}}}=\left( {\frac 32}-2{\frac \varphi \pi }%
\right) \cos \varphi +{\frac 1\pi }\left ({\frac 1{{\cos \varphi }}}-{\frac
2{{\cos {}^3\varphi }}}\right) \left( {\frac \pi 2}-\varphi -{\frac 12}\sin
(2\varphi )\right) \eqno {(A.7)}
$$
In particular, we compute the following Mac Laurin expansion:
$$
{\frac{{C(\varphi )}}{{C(0)}}}=1-2\varphi ^2+{\frac{{16}}{{3\pi }}}|\varphi
|^3+{\cal O}(\varphi ^4)\eqno {(A.8)}
$$
At $\varphi =\pi /2$ the function takes the value $-4/(3\pi )$ and vanishes
as $-(3/2)\epsilon ^2$ for $\varphi =\pi -\epsilon $.

\newpage
\centerline{Figure Captions}

\figure{
Multiple complex path used in the calculation of the
Fourier
amplitudes $a_{n}$
 with $n>0.$ Three branch points are shown.
 \label{Fig.1}}

\figure{Illustrating the scaling law (\protect\ref{scal})
  in the Fourier domain for the Elliptic GUE ensemble. The  universal
function $P_2(t)$ [ Eq.(2.14)] versus
 $t=n/ \langle |n|\rangle$ for matrices of rank
$N=71\ (\Box ), \ N=300(\triangle ),\ N=400(\bigcirc )$. Lines show the
 initial linear rise and the $t^{-4}$ tail.
   \label{Fig.2}}

\figure{Large-$n$
 decay of the Fourier coefficients of the Correlation Function, for the three
Elliptic ensembles: GOE($\bigcirc ),$ GUE($\diamond $), GSE($\triangle ).$
The Fourier coefficients have been divided by $\sqrt(C(0)$.
\label{Fig.3}}

\figure{Scaling functions
 $\Phi _\beta (X)$ versus the rescaled parameter $X$
 for Elliptic ensembles and different matrix rank $N$:
 GSE
with $N=80$ (full triangles); GSE with $N=90$ ($\bullet )$ ;
  GUE with
$N=71$ ($\bigtriangleup )$;  GUE with
 $N=300 $ ($\bigcirc $);  GUE with $N=400$ ($\Box $).
\label{Fig.4}}

\figure{The Scaling function
   $\Phi _2(X)$ for the GUE Elliptic ensemble,  matrix rank 71. The dashed line
is
the 2nd order expansion $1-2X^2;$ the solid line includes a 3d order
 term $
+1.033X^3.$ The insert illustrates the $X^{-2}$ behaviour at large
$X.$  \label{Fig.5} }

\figure{
 Variance $\sigma^2_{\ln}$ of $\ln(|a_n|^2)$ versus of $t=n/\langle n\rangle$
for the GUE Elliptic ensemble and $N=40 (\bullet),
60$$(\Box)$ and $80$$ (\triangle)$.
The horizontal dashed line corresponds to (complex) Gaussian distribution
of $a_n$
 \label {Fig.6}}
\figure{
Distribution $P(\ln r)$ of the logarithm of the mean square current, for the
GUE
Elliptic ensemble, obtained from $200$ realizations of rank $N=40$.
The dashed line represents the log-normal
distribution with $\langle \ln(r)\rangle = 1.17$
 and $\sigma^2_{\ln}=0.075$;
the solid line corresponds to the
$\chi^2_{\nu}$ distribution, with mean $c=3.34$ and
number of degrees of freedom $\nu\approx 2/\sigma^2_{\ln}$.
\label{Fig.7} }
\figure{
Illustrating the power -law  dependence
 of various ensemble averages on the scaling parameter $x=b^2/N$, for
Periodic Banded Random Matrices of rank 71,101,201.
 Average squared current $c=\langle r\rangle\ \ (\circ)$,
mean current variance  $\sigma^2_r\ \ (\Box)$,
 log-variance
$\sigma^2_{r \ln} \ \ (\diamond)$ , width of the Fourier spectrum
  $\langle |n|\rangle  \ \ (\triangle) $. Lines represent the limit dependences
that are
approached in the delocalized regime. The vertical line is somewhat arbitrarily
drawn
to represent the transition from localization to delocalization.
 \label{Fig.8}}

\figure{
Abundance of avoided crossings as a function of $\varphi$ for the PBRM
ensemble ( 150 matrices with $N=201$, $b=64$).
The histogram has been obtained from
 4452 avoided crossings occurring  inside  the open $(0,\pi)$ interval , in
which only
eigenvalues belonging in the central  half
of the spectrum were involved. The dashed line is proportional to
the average squared level velocity. The inset compares the distribution
of avoided crossing sizes (normalized to a unit mean gap) with the
Wigner distribution. \label{Fig.9}}

\figure{
CF for the PBRM ensemble in the diffusive regime ($N=201,b=64$) (inset, full
line)
and its
Fourier transform (main panel,
full circles ). The dotted line
represents the $n^{-4}$ decay. Both in the
main figure and in the inset,  dashed
lines represent the Berry -Keating
semiclassical approximation. The dotted line
in the inset represents a
semi-empirical formula proposed in ref.\cite{jz}.\label{Fig.10}}

\figure{Comparison of the scaled correlation functions $\Phi_2$ for the
Elliptic GUE ensemble with $N=71$ (circles) and $\Phi_AB$ for the
PBRM emsemble with $N=101,b=51$ (shown by a continuous line, obtained
by interpolating numerical data).\label{Fig.11}}

\figure{
the Correlation Function for the PBRM ensemble (a), and its Fourier transform
(b), for various values of the
localization parameter. Numerical data in (b) are for
matrices of rank $N=201$ and $b=101,\ 64,\ 23,\ 9$ from rightmost to
lefmost curve. All the reported data sets have the same slope, even in
the localized regime (for $b=9$, $\ln(b^2/N)\approx -1)$.
\label{Fig.12}}
\figure{The position $\varphi_0$ of the zero of the CF for the
PBRM ensemble, versus the localization parameter $x=b^2/N$ for various matrix
sizes. The dotted curve corresponds to $\varphi_0=6.55x^{1/2}/\pi$.
\label{Fig.13}}
\figure{Semilogarithmic plot of the
distribution of the coefficient $n^2a_n$ for
the PBRM ensemble, with $N=61$,$b=31$, and for $n=1$ $(\Box)$, and
$n=10$ $(\bullet)$.\label{Fig.14}}

\figure{
Average squared  velocity $\sigma^2=\langle e'(\varphi)^2 \rangle$
(normalized to
its maximum value) versus the rescaled phase $X$ , for the PBRM ensemble with
 $N=201$ and
$b=9$$(\triangle)$, $23$$(\diamond)$, $64$$(\Box)$ and $101(\bullet)$.
 In the delocalized regime numerical data
agree with the theoretical formula  of Taniguchi el al, (solid line).
 The insert shows the same data without rescaling.
\label {Fig.15}}

\end{document}